\documentclass{article}

\usepackage{amsfonts}
\usepackage{epsfig}

\def\runninghead#1#2{\pagestyle{myheadings}
\markboth{{\protect\normalsize{\quad #1}}\hfill}
{\hfill{\protect\normalsize{#2\quad}}}}
\begin{document}

\runninghead{D. Volchenkov, R. Lima} {On the convergence of
multiplicative branching processes in dynamics of  fluid flows
\ldots \hspace{1cm}}

\title{\textbf{Borel Summation Of Asymptotic Series In Dynamics Of Fluid Flows.
Diffusion Versus Bifurcations}}

\author{ D. Volchenkov \footnote{The Alexander von Humboldt Research Fellow at the
BiBoS Research Center} and  R. Lima \\
{\it  Centre de Physique Théorique (CPT), CNRS Luminy Case 907,}\\
{\it 13288 Marseille Cedex 9, France} \\
{\it E-Mails: volchen@cpt.univ-mrs.fr, lima@cpt.univ-mrs.fr}}

\date{\today}
\maketitle

\large

\begin{abstract}

The Brownian motion over the space of fluid velocity
configurations driven by the hydrodynamical equations is
considered. The Green function is computed in the form of an
asymptotic series close to the standard diffusion kernel. The high
order asymptotic coefficients are studied. Similarly to the models
of quantum field theory, the asymptotic contributions demonstrate
the factorial growth and are summated by means of Borel's
procedure. The resulting corrected diffusion spectrum has a closed
analytical form. The approach  provides a possible ground for the
optimization of existing numerical simulation algorithms and can
be used in purpose of analysis of other asymptotic series in
turbulence.

\end{abstract}
\vspace{0.5cm}

\leftline{\textbf{ PACS codes: } 47.10ad, 47.27.E-, 47.27.ef}
\vspace{0.5cm}

\leftline{\textbf{ Keywords: } Brownian motion, Asymptotic series,
Borel summation, Simulation on fluid flows }

\newpage

\section{Introduction}
\noindent

Theoretical investigations of two dimensional (2D) cross-field
transports in the operating ITER-FEAT calls for an increasing
confidence in the modelling efforts that force one to search for
the new principles of simulations.

It has been reported \cite{1,2} that  the solutions of forced
dynamical equations describing the fluid flows (such as the 3D
Fourier transformed Navier-Stokes (FNS) equation, the Burgers
equation, {\it etc}.) may be represented, point wise, as the
expected values of a branching random walk process statistic. The
statistic samples the Fourier transforms of both the initial data
and the forcing at various random frequencies as directed by the
leaves of the binary branching process, and then combines the
results together, in a multiplicative way, according to the binary
bifurcation nodes of the process. The branching random walk holds
in state $k \ge 1$, $k\in \mathbb{N}$ for an exponential time with
parameter $\nu k^2$ ( $\nu>0$ is the diffusion coefficient) then
it is replaced by no offspring with probability $1/2$, otherwise
replaced by two frequencies $j$ and $k-j$, with $j$ picked
uniformly over $[1,\dots k-1]$. The process is repeated
independently for each of the offspring sequences.

The use of finite binary tree algorithms is very plausible since
they can be easily implemented and optimized that would
drastically improve the effectiveness of large scale simulations
on fluid flows. However, it has been discovered that the
definition of the stochastic process relevant to the modelling of
complicated cross-field transports in plasma is a tough problem.
Even if such the branching multiplicative process exists, its fast
convergence is questionable. To illustrate the problem, we have
performed a preliminary simulation on the FNS. We have tried to
reproduce the well-known Kolmogorov's inertial range for the fully
developed turbulence in the 3D FNS forced in the small moments (in
the large scales). We have sampled the binary trees independently
for each of velocity components, but limit the total number of
samples just to 250 for each point $(k,t)$. For the small moments,
just a few different binary trees can be constructed and all of
them arise in samplings that fits well the fully developed
turbulence statistics (see Fig.~1). However the number of
different branching trees that could appear in simulations is huge
for $k$ large and apparently grows very fast with $k$. In this
case, the set of randomly generated 250 samples obviously does not
provide a reliable base for the turbulent statistics that is
evident from the excessively strong trembling of data worsening as
the momentum $k$ grows up. The data is presented in Fig.~1 in the
form of a box plot indicating the scattering. The central point of
a box shows the median, the lower line of a box indicates the
first quartile, and the upper line shows the third quartile. The
whiskers are extended to the maximum and minimum points. Apart
from the forced region, an inertial range appears for any
nontrivial initial condition. The strong data scattering in the
region of large moments indicates that the number of samples
should be essentially increased. It is important to mention that
the classical methods developed in \cite{1} do not provide us with
tools for the control of convergence for the stochastic
multiplicative branching processes that forces us seriously
concerns to the problem.

In the present paper, we propose an alternative method that would
optimize the simulations on fluid flows. It is closely related to
that of branching Brownian motion and uses the techniques borrowed
from the quantum field theory to control the computation accuracy.
Applications of field theory methods to the various problems of
statistical physics and critical phenomena theory have a long
history. Nowadays,they have converted into a powerful tool for the
effective studying of stochastic phenomena described by the
partial differential and integro-differential equations. The
problems of statistical physics, in their turn, had also enriched
the field theory methods as well.

We study the Green functions of equations describing the
incompressible fluid flows by constructing their functional
representations. Our approach is based on the ideas of the MSR
formalism (after Martin-Siggia-Rose) treating the classical
Langevin equations as a kind of Brownian motion. In such a
framework, the stochastic averages with respect to the Gaussian
distributed random forces stirring the system can be interpreted
as some functional averages (see Sec. 2). It is worth mentioned
that they are nontrivial even if the stochastic force is put to
zero. This fact allows modifying the initial MSR approach in a way
to apply it to the forced dynamical equations in which the
external forcing is not random. The Green functions can be
computed by means of perturbation theory in the form of Feynman
diagram series (Sec. 3).

We have demonstrated that each Feynman graph in the series equals
to an average over a forest of multiplicative branching binary
trees (of the certain topological structure) implemented in
\cite{1} that establishes the direct relation between their
methods and the field theory approach that we use (Sec. 3).
However, the diagram series converges much faster than the
 multiplicative branching process since each diagram
represents an average of branching process over all allowed
configurations of moments. We have demonstrated the advantage of
method for the stirred Burgers equation supplied with a periodic
boundary condition (Sec. 4). In this simple example, the diagrams
can be computed analytically, and the series converges very fast.
The Fourier spectrum for the Green function forms the asymptotic
series starting from the standard diffusion kernel. Fluctuations
arise due to the consequent bifurcations of media resulting in the
cascade of consequent partitions of moments,
$\mathbf{k}=\mathbf{q} + (\mathbf{k} - \mathbf{q})$. The magnitude
of relevant corrections to the diffusion spectrum tends to zero as
$t\rightarrow t_0$, and the saddle-points (instanton) analysis can
be applied to study the "large order" asymptotic contributions
(see Sec. 5). The central point of method is that a coefficient in
the asymptotic series is calculated by the Cauchy integral
$$
G^{(N)}=\frac 1{2\pi i}\oint \frac{dg}{g^{N+1}}\sum_{N=0}^{\infty}
g^N G^{(N)} = \frac 1{2\pi i}\oint \frac{dg}{g}\sum_{N=0}^{\infty}
g^N G^{(N)}\exp{(-N\log{g})}
$$
where the integration contour encircles $g=0$ in the complex
plane. If the asymptotic series $\sum_{N}G^{(N)}g^N$ is a
functional integral, but $N$ is a large number, one  can estimate
the large order coefficient $G^{(N)}$ calculating the functional
integral by the steepest descent method \cite{ZJ}. The
saddle-point equations describing the "equilibrium" state of the
hydrodynamical system with respect to the large - order
contributions into the Green function can be solved as $t\to t_0$
when the bifurcations rise the corrections to the pure relaxation
dynamics.

The calculations show that the asymptotic coefficients demonstrate
the factorial growth like the most of models in quantum field
theory. The asymptotic series for the Green function can be
summarized by means of the Borel procedure (see Sec. 5). In the
limit $t\to t_0$ the corrections to the diffusion kernel have the
closed analytical form.

The proposed approach can be broadly used in the many problems of
forced dynamics for the optimization of numerical computations and
the analysis of asymptotic series. We conclude in the last
section.

\section{Stochastic dynamics as the Brownian motion}
\noindent

It is well known that many problems in stochastic dynamics can be
treated as a generalized Brownian motion ${\left\langle {\delta
\left( {u - u\left( {x,t} \right)} \right)} \right\rangle} _{\xi}
,$ in which the classical random field indicating the position of
a particle $u(x,t)$ meets a Langevin equation
$$\dot {u}\left({x,t} \right) = Q\left( {u} \right) + \xi ,$$
where $\xi $ is the Gaussian distributed stochastic force
characterized by the correlation function $D_{\xi}  =
{\left\langle {\xi \xi} \right\rangle}.$ Here the angular brackets
${\left\langle {\ldots} \right\rangle} _{\xi}  $ denote an average
position of particle with respect to the statistics of $\xi$.
 $Q(u)$ is some nonlinear term which depends on the position
$u(x,t)$ and its spatial derivatives. Such a representation was a
key idea of the formalism \cite{Martin:1976}.

An elegant way to obtain the field theory representation of
stochastic dynamics is given by the functional integral
\begin{equation}
\label{1}
 {\left\langle {\delta \left( {u - u\left( {x,t}
\right)} \right)} \right\rangle} _{\xi}  \equiv \int {D\xi \exp
{\rm T}{\rm r}\left( { - {\frac{{1}}{{2}}}\xi D_{\xi}  \xi}
\right)\delta \left( {u - u\left( {x,t} \right)} \right)}
\end{equation}
where the Tr-operation means the integration $\int {dx dt} $ and
the summation over the discrete indices. The instantaneous
positions $u\left( {x,t} \right)$ meet the dynamical equation that
can be taken into account by the change of variables
\begin{equation}\label{2}
 \delta \left( {u - u\left( {x,t} \right)} \right) \to \delta
\left( {\dot {u}\left( {x,t} \right) - Q\left( {u} \right) - \xi}
\right)
\end{equation}
should the solution of dynamic equation exists and is unique. The
use of integral representation for the $\delta - $function in
(\ref{1}) transforms it into
\begin{equation}\label{3}
 \int {D\xi DuDu'\exp {\rm T}{\rm r}\left( { - {\frac{{1}}{{2}}}\xi
D_{\xi} \xi - \dot {u}u' + u'Q\left( {u} \right) + u'\xi} \right)}
\det M,
\end{equation}
in which $u'\left( {x,t} \right)$ is the auxiliary field that is
not inherent to the original model, but appears since we treat its
dynamics as a Brownian motion. The Jacobian $\det M$ relevant to
the change of variables (\ref{2}) is discussed later.

The Gaussian functional integral with respect to the stochastic
force $\xi$ in (\ref{3}) is calculated
\begin{equation}
\label{4} \int {DuDu'} \exp \left( {S\left( {u,\eta}  \right)}
\right)\det M, \quad S(u,u') = {\rm T}{\rm r}{\left[ { -
{\frac{{1}}{{2}}}u'D_{\xi}  u' - u'\dot {u} + u'Q\left( {u}
\right)} \right]}.
\end{equation}
By means of that all configurations of $\xi $ compatible with the
statistics are taken into account. The integral (\ref{3})
identifies the statistical averages ${\left\langle {\ldots}
\right\rangle} _{\xi} $ with the functional averages of weight
$\exp S$. The formal convergence requires the field $u$ to be real
and the field ${u}'$ to be purely imaginary.

The functional averages in (\ref{4}) can be represented by the
standard Feynman diagram series exactly matching (diagram by
diagram) the usual diagram series found by the direct iterations
of the Langevin equation averaged with respect to the random force
This fact justifies the use of functional integrals in stochastic
dynamics at least as a convenient language for the proper diagram
expansions.

The Jacobian $\det M$ in (\ref{3}) depends upon the nonlinearity
$Q\left( {u} \right)$. If $Q\left( {u} \right)$ does not depend
upon the time derivatives, all diagrams for $\det M$ are the
cycles of retarded lines $ \overleftarrow{\Delta}\propto \theta
\left( {t - t'} \right)$ and equal to zero excepting for the very
first term,
\begin{equation}
\label{6} \det M = const \cdot \exp {\rm Tr}\Delta ,
\end{equation}
Following \cite{Adzhemyan:1999}, in the paper, we take the
convention for the Heaviside function of zero argument,
$\theta(0)=0$, so that $\det M = const.$

Below, we consider the field theory formulations of equations
describing the fluid flows. We suppose that the fluid is
incompressible, ${\rm d}{\rm i}{\rm v}{\ } u = 0$. In the Fourier
space, the relevant transversal (rotational) component of $u$ can
be allocated by means of the transversal projector,
\[
P_{ \bot}  \left( {k} \right) = \delta _{ij} - {\frac{{k_{i} k_{j}
}}{{k^{2}}}}{\rm .}
\]
The transversal projection of Navier-Stocks equation (NS) lacks of
the pressure gradient and the longitudinal component of nonlinear
term, ${\left[ {\left( {u\partial}  \right)u} \right]}_{\parallel}
$. Any solution of it corresponds to some vortex of momentum $k$.
The MSR action functional for the transversal NS is
\begin{equation}
\label{7} S\left( {u,u'} \right) = {\rm T}{\rm r}{\left[ { -
{\frac{{1}}{{2}}}u'D_{\xi } u' - u'\dot {u} + \nu u'\Delta u -
u'V\left( {uu} \right)} \right]}
\end{equation}
where $V$ is the differential operator $V_{ijs} = {{i\left( {k_{j}
\delta _{is} + k_{s} \delta _{ij}} \right)} \mathord{\left/
{\vphantom {{i\left( {k_{j} \delta _{is} + k_{s} \delta _{ij}}
\right)} {2}}} \right. \kern-\nulldelimiterspace} {2}}$ (in the
Fourier space). The field theory with the action functional
(\ref{7}) has been discussed in details in \cite{Adzhemyan:1999}.
The functional averages computed with respect to the statistical
weight $\exp{\ } S$ can be expanded into the series of Feynman
diagrams drawn with the interaction vertex
\begin{figure}[h]
 \noindent
 \begin{center}
 \begin{minipage}[h]{0.46\linewidth}
 \epsfig{file=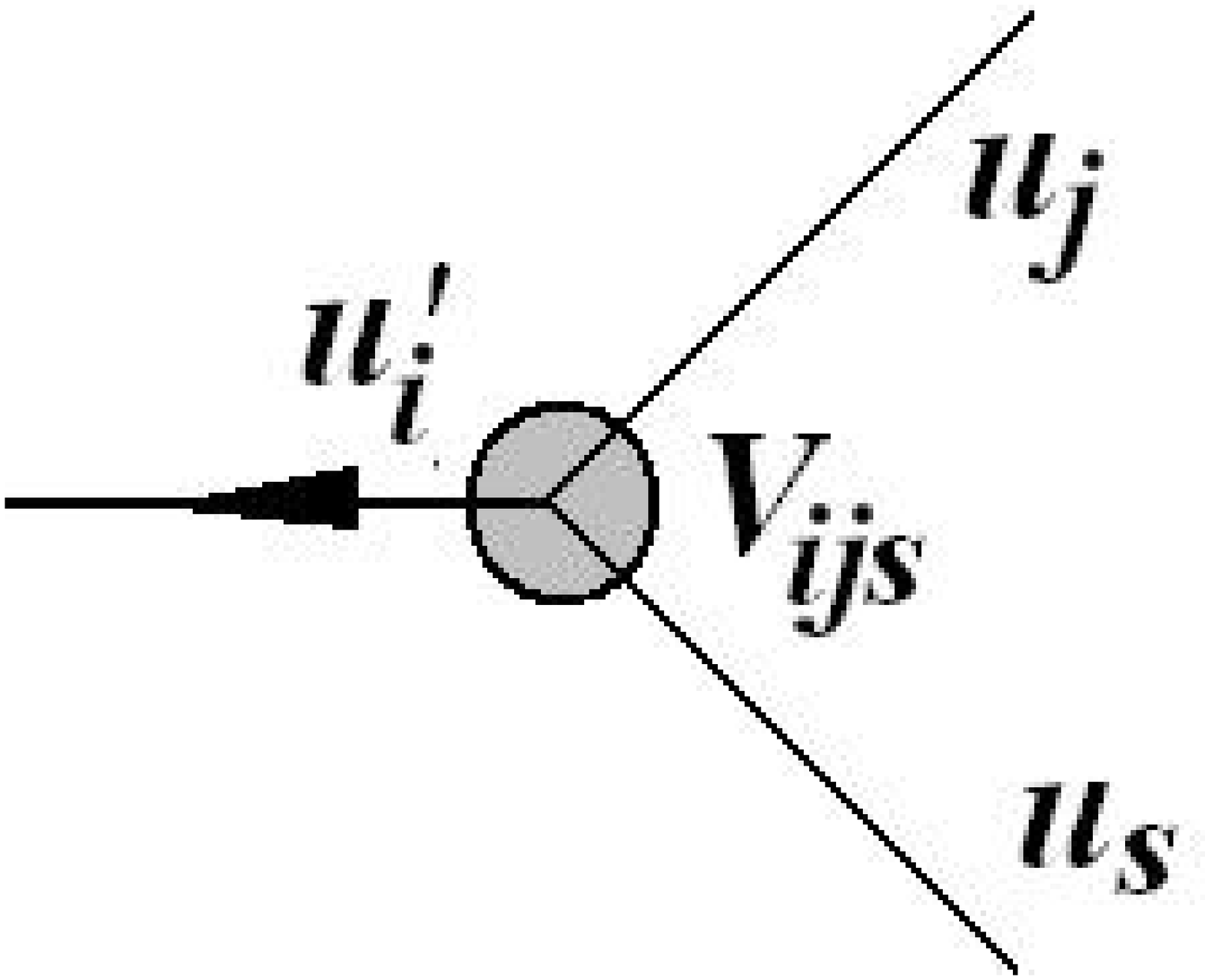,angle=0,width=3.9cm,height=2.7cm}
 \end{minipage}
\end{center}
\end{figure}
in which the spatial derivative $V(k)$ is encircled and the arrow
denotes the $u'$-tail, and two propagators (lines) which have the
following analytical representations (in the Fourier space)
\begin{equation}\label{8}
 \Delta_{uu'} ={\frac{{P_{ \bot} } }{{\left( { - i\omega +
\nu k^{2}} \right)}}},\quad \Delta_{uu} = {\frac{{P_{ \bot}
D_{\xi} } }{{\left( {\omega ^{2} + \nu ^{2}k^{2}} \right)}}}
\end{equation}
The inverse  Fourier transform of $\Delta_{uu'} $ shows that it is
retarded, $\Delta_{uu'} \propto \theta(t-t_0).$ Graphs for the
Green functions in the theory with the action functional (\ref{7})
exactly match the well-known diagram expansion of Wyld \cite{W}
developed by the direct  iterations of stochastically driven NS
with respect to its nonlinear term with the consequent averaging
over the Gaussian distributed random force.

One can hardly formulate a general method converting a problem of
stochastic dynamics into a field theory (there are many, \cite{D}
) but once it is done, it allows for applying the powerful
techniques within the well controlled approximations.

\section{ Branching representations to the Green function of Navier-Stokes  }
\noindent

Many dynamical systems are driven by the non-random external
forces. It is worth to mention that if one assumes  $\left| \xi
\right|\rightarrow 0$ (and consequently $D_\xi\rightarrow 0$) in
the action functional (\ref{7}), it  is retained, but then the
diagram series is trivial since the velocity field propagator
vanishes, $\Delta_{uu} = 0$. However, the diagram series would be
recovered by means of regular external forcing.

For instance, let us consider the Cauchy problem for the NS
equation,
\begin{equation}
\label{9} \dot {u}_{j} + \partial _{i}
\left( {u_{j} u_{i}}  \right) = \nu \Delta u_{j} + \delta \left(
{x - x_{0}}  \right)\delta \left( {t - t_{0}} \right){\rm ,}
\end{equation}
supplied with the localized integrable initial condition $u_{0} =
u\left( {0,x} \right)$ that means the external force $f\left(
{x,t} \right) = \delta \left( {x - x_{0}}  \right)\delta \left( {t
- t_{0}}  \right)$ acting on the system. The relevant action
functional
\begin{equation}
\label{10} S\left( {u,u'} \right) = {\rm T}{\rm r}{\left[ { -
u'\dot {u} - g\nu u'V\left( {uu} \right) + \nu u'\Delta u}
\right]} + u'\left( {x_{0} ,t_{0}} \right)
\end{equation}
includes the ultra-local interaction term $u'\left( {x_{0} ,t_{0}}
\right)$. To obtain a formal expansion parameter in the
perturbation theory for (\ref{10}), we have inserted the coupling
constant $g\nu \equiv 1$ in front of the interaction term. Despite
its physical dimension
 ${\left[ {g} \right]} = - {\left[ {\nu}  \right]}{\rm ,}$ its conventional
dimension would be different. The appearance of ultra-local terms
in a field theory had been discussed in \cite{Symanzik:1981}. The
basic symmetry of (10) is the Galilean invariance,
\begin{equation}
\label{11} u_{a} \left( {x,t} \right) \to u\left( {x + X_{a}
\left( {t} \right),t} \right) - a\left( {t} \right),{\rm} {\rm}
u'\left( {x,t} \right) \to u'\left( {x + X_{a} \left( {t}
\right),t} \right)
\end{equation}
where $a\left( {t} \right)$ is an arbitrary function of time
decaying rapidly as ${\left| {t} \right|} \to \infty ,$ and $X_{a}
\left( {t} \right) = {\int\limits_{0}^{t} {a\left( {\tau}
\right)}} d\tau $. The transformations (\ref{11}) define the set
of orbits in the configuration space along which the functional
averages do not change. It follows that the functional integral
itself is proportional to the volume of such orbits. The Green
function $G(x,t;x_{0} ,t_{0} )$ for the Cauchy problem (\ref{9})
can be computed as the functional average,
\begin{equation}\label{12}
G(x,t;x_{0} ,t_{0} ) = {\left\langle {u} \right\rangle}  =
{\frac{{\int\!\!\!\int {u(x,t)\exp S(u,u')DuDu'}}
}{{\int\!\!\!\int {\exp \left( {S_{0}}  \right)DuDu'}}
}},{\rm}\quad {\rm} S_{0} = Tr{\left[ { - u'\dot {u} + \nu
u'\Delta u} \right]},
\end{equation}
in which $S$ is the action functional (\ref{10}). The above
functional average is interpreted as an infinite diagram series,
\begin{equation}
\label{13} G(x,t;x_{0} ,t_{0} ) =
\end{equation}
\begin{flushleft}
 \begin{minipage}[h]{0.76\linewidth}
 \epsfig{file=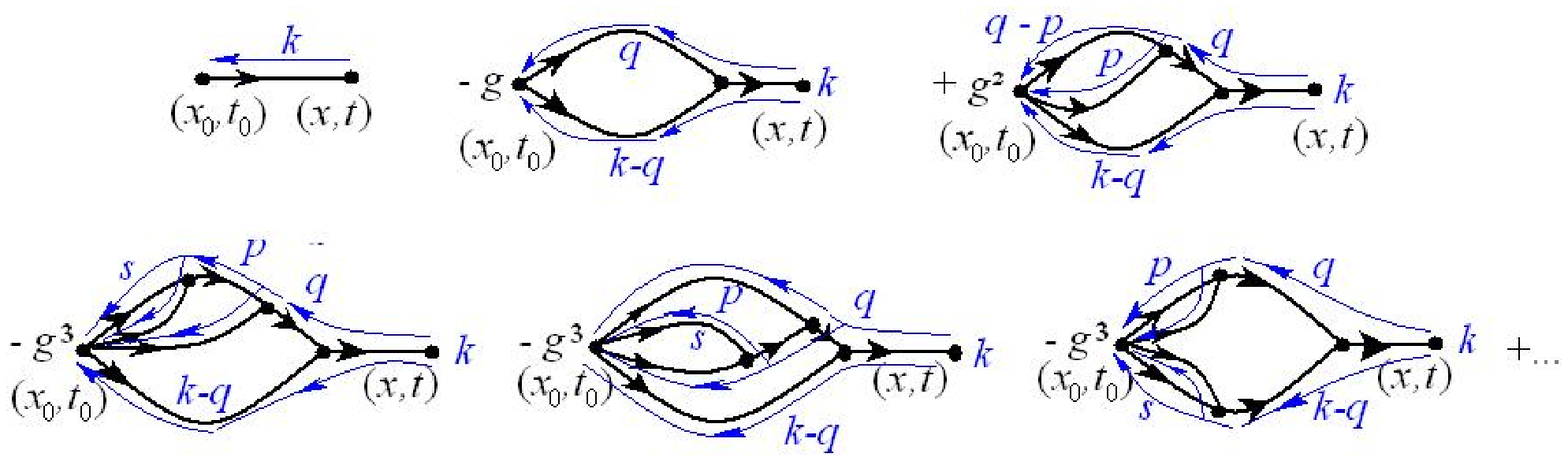,angle=0,height=4.5cm,width=14cm}
 \end{minipage}
\end{flushleft}
where the arrows on the bold lines indicate the time direction;
the thin arrows display the momentum flux traversing the
correspondent branch of a graph. The moments are to be conserved
at each node of any graph. In comparison to the standard Wyld
diagrams the graphs sketched in (\ref{13}) contain the local
interaction vertex with any number of $u'$-tails

\begin{figure}[h]
 \noindent
 \begin{center}
 \begin{minipage}[h]{0.46\linewidth}
 \epsfig{file=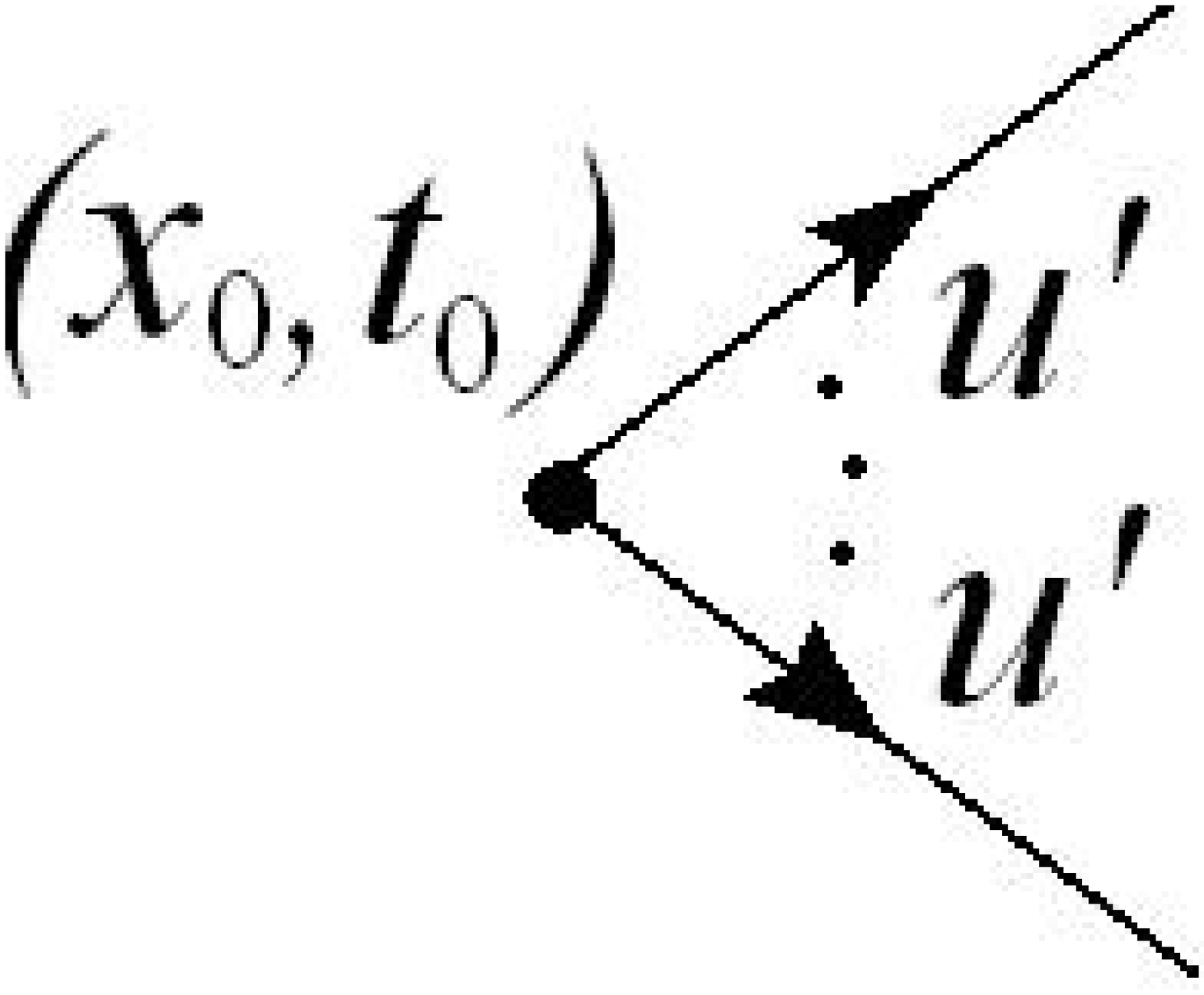,angle=0,width=3.9cm,height=2.7cm}
 \end{minipage}
\end{center}
\end{figure}

that recovers the diagrams even if the only line
$\overleftarrow{\Delta}_{uu'}$ is retarded. The similar diagrams
have been derived for the problem of nonlinear diffusion
\cite{Antonov:2002}, however, in contrast to them, each vertex in
the diagram series (\ref{13}) contains the differential operator
$V$. The coupling constant $g \equiv {{1} \mathord{\left/
{\vphantom {{1} {\nu} }} \right. \kern-\nulldelimiterspace} {\nu}
}$ plays the role of an expansion parameter, so that any graph has
the amplitude factor $g^{\ell}$ where $\ell$ is the number of
loops presented in it.

The diagram expansion for the Green function would have a definite
physical meaning if it converges. The standard analysis of
ultraviolet divergences of graphs is based on the counting of
relevant canonical dimensions. Dynamical models have two scales,
the time scale $T$ and the length scale $L$, consequently the
physical dimension of any quantity $F$ can be defined as ${\left[
{F} \right]} = L^{ - d_{F}^{k}} T^{ - d_{F}^{\omega} } ,$ in which
$d_{F}^{k} $ and $d_{F}^{\omega}  $ are the momentum and frequency
dimensions of $F$. In diffusion models, these dimensions are
always related to each other since
 $\partial _{t} \sim \partial^2_x $ in the diffusion equation that allows
us to introduce a combined canonical dimension, $d_{F} =
d_{F}^{\omega}  + 2d_{F}^{k} $. One can check out that each term
in (\ref{10}) is dimensionless if the following relations hold:
$d_{u'} = 0,d_{u} = d,d_{\nu}  = - 2 + 2 \cdot 1 = 0,$ and $d_{g}
= 2 - \left( {d + 1} \right).$ The field theory (\ref{10}) is
logarithmic (the conventional dimension of the coupling constant
$d_{g} = 0$) for the Burgers equation $\left( {d = 1} \right)$,
while in two dimensions $d_{g} = - 1{\rm ,}$ and $d_{g} = - 2$ for
$d = 3$ (the NS equation). Thus, in the infrared region (small
moments, large scales) the diagram series in $g$ define just the
corrections to the diffusion kernel as $d \ge 2$. However, in the
case of Burgers equation, all diagrams look equally essential in
large scales.

The diagrams diverge in the ultraviolet region (large moments,
small scales) if their canonical dimension
\[
d_{\Gamma}  = - d_{u} N_{u} - d_{u'} N_{u'} \ge 0
\]
where $N_{u} $ and $N_{u'} $ are the numbers of corresponding
external legs in the graph $\Gamma$. For the Green function
(\ref{12}), we have $N_{u'} = 0$ and $N_{u} = 1$. Therefore, there
is no any ultraviolet divergent graph in the diagram series
(\ref{13}). For the first glance, it seems that any graph having
no external $u - $legs $\left( {N_{u} = 0} \right)$ and any number
of auxiliary fields $u'$ $\left( {N_{u'} > 0} \right)$ should
diverge since $d_{u'} = 0$. However, such a graph is also
convergent in small scales because of the derivatives $V(k)$ which
are always taken outside the graph onto the external $u' - $legs
that effectively reduces its canonical dimension to ${d}'_{\Gamma}
= d_{\Gamma}  - N_{u'} < 0{\rm .}$ Therefore, the theory
(\ref{10}) has no ultraviolet divergences and need not to be
renormalized.

Each Feynman diagram in (\ref{13}) series corresponds to a certain
gyro-dynamical process contributing into $G(x,t;x_{0} ,t_{0} )$.
For the convenience of readers, we give the parallel
interpretation of diagrams in terms of multiplicative branching
stochastic processes discussed in \cite{1,2,Ossiander:2000}. The
very first diagram is presents the solution of diffusion equation
(in $d$-dimensional space),
\begin{equation}
\label{14} \Delta \left( {x - x_{0} ,t - t_{0}}  \right) =
{\frac{{\exp \left( { - {{\left( {x - x_{0}}  \right)^{2}}
\mathord{\left/ {\vphantom {{\left( {x - x_{0}}  \right)^{2}}
{4\nu \left( {t - t_{0}}  \right)}}} \right.
\kern-\nulldelimiterspace} {4\nu \left( {t - t_{0}}  \right)}}}
\right)}}{{\left( {4\pi \nu \left( {t - t_{0}}  \right)}
\right)^{{{d} \mathord{\left/ {\vphantom {{d} {2}}} \right.
\kern-\nulldelimiterspace} {2}}}}}}.
\end{equation}
It describes the simple viscous dissipation of a vortex with no
bifurcations. The second graph in (\ref{13}) corresponds to a
 process of twofold splitting. Under the spatial Fourier
transformation, it is equivalent to the following analytic
expression:
\begin{equation}
\label{15} \Gamma _{1} \left( {{\rm {\bf k}},t} \right) = -
g{\int\limits_{0}^{\infty} {d{t}'}} {\rm} V\left( {{\rm {\bf k}}}
\right)\Delta \left( {{\rm {\bf k}},t - {t}'} \right)\int
{{\frac{{d{\rm {\bf q}}}}{{\left( {2\pi} \right)^{d}}}}} \Delta
\left( {{\rm {\bf q}},{t}'} \right)u_{0} \left( {{\rm {\bf q}}}
\right)\Delta \left( {{\rm {\bf k}} - {\rm {\bf q}},{t}'}
\right)u_{0} \left( {{\rm {\bf k}} - {\rm {\bf q}}} \right),
\end{equation}
where $\Delta \left( {{\rm {\bf k}},t} \right)$ is the spatial
Fourier transform of the diffusion kernel (14) and $u_{0} \left(
{{\rm {\bf k}}} \right)$ is the Fourier spectrum of initial
condition. It is worth to mention that the diagram expansion for
the Green function can be also discussed for the function defined
a the finite domain supplied with a periodic boundary condition.
In the latter case, it has a discrete set of harmonics, and the
integral (\ref{15}) turns into sums.

The time integration in (\ref{15}) can be performed easily,
\begin{equation}
\label{16} \Gamma _{1} = - gV\left( {{\rm {\bf k}}} \right)\exp
\left( { - \nu k^{2}\left( {t - t_{0}}  \right)}
\right){\int_{{\rm {\bf q}} \cdot {\rm {\bf k}} > q^{2}}
{{\frac{{d{\rm {\bf q}}}}{{\left( {2\pi} \right)^{d}}}}{\left[
{u_{0} \left( {{\rm {\bf q}}} \right)u_{0} \left( {{\rm {\bf k}} -
{\rm {\bf q}}} \right){\frac{{\exp ( - 2\nu \left( {{\rm {\bf q}}
\cdot {\rm {\bf k}} - q^{2}} \right)t_{0} )}}{{2\nu \left( {{\rm
{\bf q}} \cdot {\rm {\bf k}} - q^{2}} \right)}}}} \right]}}} .
\end{equation}
The singularities in $\Gamma _{1} $ appear at ${\rm {\bf q}} = 0$
and ${\rm {\bf q}} = {\rm {\bf k}}$, when the vortex does not
bifurcate. The remaining momentum integral in (\ref{16}) can be
interpreted as an expectation value,
\begin{equation}
\label{17} \wp _{1} \left( {{\rm {\bf k}}} \right) =
{\int\limits_{t_{0}} ^{\infty} {d\tau} } {\int_{} {{\frac{{d{\rm
{\bf q}}}}{{\left( {2\pi} \right)^{d}}}}{\left[ {u_{0} \left(
{{\rm {\bf q}}} \right)u_{0} \left( {{\rm {\bf k}} - {\rm {\bf
q}}} \right)\exp \left( { - 2\nu \left( {{\rm {\bf q}} \cdot {\rm
{\bf k}} - q^{2}} \right)\tau}  \right)} \right]}}} ,
\end{equation}
over the Poisson process of vortex bifurcation at momentum ${\rm
{\bf k}}$. Consider the statistical ensemble of the pitchfork
bifurcations (see Fig.~2.a) relevant to the set of possible
branching random walks discussed in \cite{1}.

For each simple tree, the processes start at the vortex with the
fixed momentum $k \ge 1$. The branching random walk holds in state
$k$ for an exponential time characterized by the parameter $\nu
k^{2}$. When clock rings, the vortex bifurcates into two offspring
with moments $j \in {\left[ {1,k - 1} \right]}$ and $k - j$. They
exist for the exponential times accordingly to their moments and
then dissipate (that is figured by the crosses in Fig.~2.a). In
accordance to the rules of \cite{1}, at each vertex of momentum
$q$ which dies, one attaches the initial condition $u_{0} \left(
{q} \right)$. The solutions of the fluid flows equations (defined
in a finite domain with periodic
 $L_{1} $ initial data) are the expectations of a multiplicative
functional that is constituted by the recursive application of the
following rule to each node: $ - i$ times the product of the pair
of values attached to the corresponding pair of input nodes. Each
pitchfork bifurcation tree shown in Fig.~2.a presents one of
possible primitive realizations of the recursive multiplicative
algorithm from \cite{1}. The analytical expression (\ref{15}) for
the second diagram sums up the contributions coming from all
pitchfork trees in the forest shown in Fig.~2.

The two-loop diagram in (\ref{13}) represents the forest of
multiplicative branching random walks with the double pitchfork
bifurcations schematically shown on Fig.2.b. The graph corresponds
to the following analytical expression:
\begin{equation}
\label{18} g^{2}V\left( {{\rm {\bf k}}} \right)\exp \left( { - \nu
k^{2}\left( {t - t_{0}}  \right)} \right)\times \wp _{2} \left(
{{\rm {\bf k}}} \right)
\end{equation}
where the functional
\[
\wp _{2} \left( {{\rm {\bf k}}} \right) = {\int\limits_{t_{0}}
^{\infty} {d{t}'}} \int {{\frac{{d{\rm {\bf p}}}}{{\left( {2\pi}
\right)^{d}}}}} u_{0} \left( {{\rm {\bf k}} - {\rm {\bf p}}}
\right)V\left( {{\rm {\bf p}}} \right)\wp _{1} \left( {{\rm {\bf
p}}} \right)\exp \left( { - 2\nu \left( {{\rm {\bf p}} \cdot {\rm
{\bf k}} - p^{2}} \right){t}'} \right)
\]
and $\wp _{1} \left( {{\rm {\bf k}}} \right)$ is defined in (17).
The diagrams can be computed provided an integrable initial
condition
\begin{equation}
\label{19} u_{0} \left( {k} \right) = {\sum\limits_{m}
{{\frac{{c_{m}} }{{k^{m}}}}} },{\rm} {\rm}
\end{equation}
is given. In (\ref{19}), the sum is taken over $m$ odd if the
space dimension $d$ is even, and over $m$ even otherwise. For each
term in (\ref{19}), one can compute the functional $\wp _{\ell}
\left( {k} \right)$ of arbitrary order $\ell $,
\begin{equation}
\label{20} \wp _{\ell}  \left( {k} \right) = {\frac{{c_{m}}
}{{\left( {4\pi} \right)^{{{d\ell}  \mathord{\left/ {\vphantom
{{d\ell}  {2}}} \right. \kern-\nulldelimiterspace}
{2}}}}}}k^{\gamma _{\ell} } {\left[ {{\frac{{\Gamma \left(
{{\frac{{d}}{{2}}} - {\frac{{m + 1}}{{2}}}} \right)}}{{\Gamma
\left( {{\frac{{m + 1}}{{2}}}} \right)}}}} \right]}^{\left( {\ell
+ 1} \right)}{\mathop {\prod} \nolimits_{s = 1}^{\ell }} \left(
{{\frac{{\Gamma \left( {{\frac{{d}}{{2}}} - {\frac{{\gamma _{s}
}}{{2}}}} \right)}}{{\Gamma \left( {{\frac{{\gamma _{s}} }{{2}}}}
\right)}}}{\frac{{\Gamma \left( {{\frac{{\gamma _{s} + m + 1 -
d}}{{2}}}} \right)}}{{\Gamma \left( {d - {\frac{{\gamma _{s} + m +
1}}{{2}}}} \right)}}}} \right),
\end{equation}
where $\gamma _{\ell}  = \ell d - \left( {\ell + 1} \right)\left(
{m + 1} \right)$. Accounting for several terms in (\ref{19}) makes
the analytical computations very difficult, nevertheless they can
be performed numerically.

\section{One-dimensional diffusion with bifurcations}
\noindent

In the present section we study the correction to the 1D diffusion
spectrum arisen due to the Poisson stochastic process of momentum
bifurcations. For the sake of simplicity, we consider the one
dimensional diffusion problem supplied with the initial condition
in the form of an infinite power series,
\begin{equation}
\label{21} u_{0} \left( {k} \right) = {\sum\nolimits_{m} {k^{ -
m}}} ,k > 0
\end{equation}
that corresponds to the function $u_{0} \left( {x} \right) = -
{{\sin \left( {x} \right)} \mathord{\left/ {\vphantom {{\sin
\left( {x} \right)} {2}}} \right. \kern-\nulldelimiterspace} {2}}$
in the real space. The $k$-spectrum of the unperturbed diffusion
kernel at $t = 5,$ $\nu = 0.005$ is shown by the solid line on
Fig.~3. The dash line presents the $G\left( {k} \right)$-spectrum
accounting for the corrections due to the pitchfork bifurcations
with
\begin{equation}
\label{22} \wp _{1} \left( {k} \right) = {\frac{{1}}{{4\pi
k}}}\left( {\cos \left( {k} \right)\left( {{\rm C}{\rm i}\left(
{2k} \right) + {\rm C}{\rm i}\left( { - 2k} \right) - \log \left(
{ - k^{2}} \right) - 2\log 2 - 2\gamma}  \right) + 2{\rm S}{\rm
i}\left( {2k} \right)\sin \left( {k} \right)} \right)
\end{equation}
where ${\rm C}{\rm i}\left( {k} \right)$ is the cosine integral,
${\rm S}{\rm i}\left( {k} \right)$ is the sine integral, $\gamma $
is the Euler's constant. The $G\left( {k} \right)$- spectrum
accounting for the corrections due to both the second and the
third diagrams in (\ref{13}) is displayed on Fig.~3 by the
dash-dot line. We are not assured that the expansion parameter $g$
 is small, and the expansion (\ref{13}) is
the asymptotic series. The fact that the latter spectrums
 coincide gives us only the indirect evidence in favor
 of relatively fast convergence rate of
 diagram contributions in the series. We discuss this problem
 in more details in the
forthcoming section.

Bifurcation of vortexes is the Poisson stochastic process
developing with time. On the outlined graphs, we have displayed
the exponential decay of $G$-spectrums at given momentum $k$. In
the real space, the Green function $G$ comprises of several
typical patterns of sizes $ \propto 1/k_m$ where $k_{m} $ are the
local maximums of the spectrum presented on Fig.~3.

\section{The large order asymptotic analysis for the
Green function: instanton approach and Borel summation } \noindent

We have mentioned in the previous section that the expansion
parameter $g$ in (\ref{13}) is not small. To get the information
on the convergence of asymptotic series, we study the asymptotic
behavior of large order coefficients $G^{(N)}$ in the diagram
series
$$
G(g)=\sum_{k=1}^N G^{(N)}g^N,
$$
as $N\to \infty,$ by the asymptotic calculation of the Cauchy
integral,
\begin{equation}\label{23}
G^{(N)}={\frac{{1}}{{2\pi i}}}\oint {{\frac{{dg}}{{g}}}} G\left(
{x,t;x_{0} ,t_{0}} \right)\exp \left( { - N\log g} \right),
\end{equation}
in which $G(x,t;x_0,t_0)$ is the functional integral (\ref{12}).
The contour of integration in (\ref{23}) embraces the point $g=0$
in the complex plain. We compute the functional integral by the
steepest descent method supposing that $N$ is large (instanton
method). Instanton approach has been applied to the various
problems of stochastic dynamics
\cite{Gurarie:1996,Volchenkov:2001,Andreanov:2006,Honkonen:2005,Honkonen:2006}.

Following the traditional instanton analysis, we perform the
rescaling of variables in the action functional (\ref{10}) in
order to extract their dependence of $N$,
\begin{equation}
\label{24} u \to u\sqrt {N} ,{\rm} {\rm} {u}' \to {u}'\sqrt {N}
,{\rm} {\rm} g \to g/N, {\rm} {\rm} \nu \to N\nu,{\rm} {\rm} x\to
x \sqrt{N}.
\end{equation}
This dilatation keeps the action functional (\ref{10}) unchanged,
thus each term acquires the multiplier $N$ and then formally gets
the same order, as the $\log g$ in (\ref{23}). The corresponding
Jacobians from the numerator and the denominator of (\ref{12})
cancel. The saddle point equations are
\begin{equation}\label{25}
  {\dot {u}}' + g\nu uV\left( {{u}'} \right) + \nu \Delta {u}' = 0
\end{equation}
\begin{equation}\label{26}
  \dot {u} + g\nu V\left( {uu} \right) - \nu \Delta u - \delta \left( {x -
x_{0}}  \right)\delta \left( {t - t_{0}}  \right) = 0
\end{equation}
\begin{equation}\label{27}
  \nu {u}'V\left( {uu} \right) = {\frac{{1}}{{g}}}
\end{equation}
The first equation (\ref{25}) recovers the original Cauchy
problem. The additional equations occur within the framework of
our approach since we analyze the dynamics as a kind of Brownian
motion. Eq.~(\ref{26}) for the auxiliary field is characterized by
the negative viscosity, and therefore $u'(t>t_0)=0$. The last
equation determines the saddle-point value of $g$ that allows to
exclude the interaction from the saddle-point equations,
\begin{equation}\label{28}
  \begin{array}{c}
   u' \Lambda u=-1 + u'\delta(\mathbf{x}-\mathbf{x}_0)\delta(t-t_0), \\
   u \Lambda^{*}u'=1, \
  \end{array}
\end{equation}
in which we have introduced the diffusion kernel
$\Lambda=\partial_t+\nu k^2,$ and its complex conjugate (in the
Fourier space) $\Lambda^{*}=\partial_t-\nu k^2$. It is interesting
to note that the possible solutions of (\ref{28}) should meet the
anti-commutation relation,
$$
u'\Lambda u+u\Lambda^{*}u'= u'(x_0,t_0)
\delta(\mathbf{x}-\mathbf{x}_0)\delta(t-t_0).
$$
Bifurcations of vortexes arisen due to the nonlinearity of
hydrodynamic equations do not conclude into a critical regime in
the model (\ref{10}), and therefore the time spectrum in the
nonlinear model is the same as for the free diffusion equations,
\textit{T$\sim $L}$^{{\rm 2}}$, that is the reason for the
branching process to be Poisson distributed with the
characteristic time 1/\textit{$\nu $k}$^{{\rm 2}}$. One can see
that the saddle-point configurations which enjoy (\ref{28}) should
not depend upon the Poisson branching, therefore, the solutions
could exist before the bifurcations commence that is as $t\to
t_0.$ Taking into account that
\[
\delta \left( {x} \right) = {\mathop {\lim} \limits_{\varepsilon
\to 0} }{\frac{{\varepsilon} }{{\pi \left( {x^{2} + \varepsilon
^{2}} \right)}}}{\rm ,}
\]
one can find that in the limit $t \to t_0$ (\ref{28}) is satisfied
by the following radially symmetric solutions,
\begin{equation}\label{29}
  u\left( {{\rm {\bf r}},t} \right) = \sqrt {\left( {{\rm {\bf r}} - {\rm {\bf
r}}_{0}}  \right)^{2} + \left( {t - t_{0}}  \right)^{2}} ,{\rm}
{\rm }{u}'\left( {{\rm {\bf r}},t} \right) = {\frac{{\theta \left(
{t_{0} - t} \right)}}{{t - t_{0}} }}u\left( {{\rm {\bf r}},t}
\right),
\end{equation}
\begin{equation}\label{30}
u\left( {{\rm {\bf r}},t} \right) = {\frac{{\theta \left( {t -
t_{0}} \right)}}{{\pi} }}\arctan \left( {{\frac{{{\left| {{\rm
{\bf r}} - {\rm {\bf r}}_{0}}  \right|}}}{{2\nu \left( {t - t_{0}}
\right)}}}} \right),{\rm }{\rm} {u}'\left( {{\rm {\bf r}},t}
\right) = - {\frac{{\pi} }{{2\nu }}}{\frac{{\left( {4\nu
^{2}\left( {t - t_{0}}  \right)^{2} + \left( {{\rm {\bf r}} - {\rm
{\bf r}}_{0}}  \right)^{2}} \right)}}{{{\left| {{\rm {\bf r}} -
{\rm {\bf r}}_{0}}  \right|}}}},
\end{equation}
\begin{equation}\label{31}
u\left( {{\rm {\bf r}},t} \right) = {\frac{{\theta \left( {t -
t_{0}} \right)}}{{\pi} }}\arctan \left( {{\frac{{2\nu \left( {t -
t_{0}} \right)}}{{{\left| {{\rm {\bf r}} - {\rm {\bf r}}_{0}}
\right|}}}}} \right),{\rm} {\rm} {u}'\left( {{\rm {\bf r}},t}
\right) = {\frac{{\pi }}{{2\nu} }}{\frac{{\left( {4\nu ^{2}\left(
{t - t_{0}}  \right)^{2} + \left( {{\rm {\bf r}} - {\rm {\bf
r}}_{0}}  \right)^{2}} \right)}}{{{\left| {{\rm {\bf r}} - {\rm
{\bf r}}_{0}}  \right|}}}},
\end{equation}
The solutions (\ref{29}-\ref{31}) are displayed on Figs.~4-5.

 The auxiliary field $u'$ in (\ref{29}) has a pole at
$t=t_0$, and then from (\ref{27}) it follows that $g_*=0$ and
therefore lays inside the integration contour in (\ref{23}). In
contrast to it, in (\ref{31}), $u\to 0 $ as $t\to t_0$ and
consequently, $g_*\to\infty$ that is definitely outside the
integration contour. Eq.~(\ref{30}) is a subtle point since
$u(r,t=t_0)=\theta(0)/2,$ and the position of saddle-point
configuration charge $g_*$ depends upon the conventional value
$\theta(0).$ While estimating the functional Jacobian in
(\ref{6}), we had assumed that $\theta(0)=0$. Following such a
convention, one can conclude that in this case it is also $g_*\to
\infty$, so that being interested in the large order asymptotic
behavior of $G^{(N)}$ we do not need to take the solution
(\ref{30}) into account. Even if one takes $\theta(0)$ as finite,
then $0<g_*<\infty,$ and it is always possible to distort the
integration contour in such a way to avoid $g_*$ to be encircled.
Therefore, (\ref{29}) is the only solution we need, and below we
use
$$
g_*\equiv \lim_{\delta t \to 0} g \simeq \frac{\delta t}{\nu
r^{2}}.
$$
Fields $u$, $u',$ and the parameter $g$ fluctuate around their
saddle-point values $u_*,$ $u'_*$, and $g_*$. By means of the
standard shift of variables, $u=u_*+\delta u,$ $\delta
u'=u'_*+\delta u',$ $g=g_*+\delta g$, one makes them fluctuate
around zero, so that $\delta u (\infty)=0$, $\delta u'(\infty)=0$.
Moreover, $\mathrm{Tr}{\ } \delta u =\mathrm{Tr}{\ } \delta u' =0$
due to the isotropy of fluid, and therefore these fluctuations are
posses the central symmetry, $\delta u=\delta u (r,t)$, $\delta
u'=\delta u' (r,t)$, the same as the saddle point configuration.

The contour of integration on the variable $\delta g$ now passes
through the origin point, and is directed there oppositely to the
imaginary axis. The integral on $\delta g$ is conducted now on a
rectilinear contour in complex plane $(i\infty, -i\infty)$ (in
accordance to the standard transformation of a contour in the
method of the steepest descent). At the turn of the integration
contour $\delta g \to i\delta g$,  the multiplier $(-i)$ appears
so the result $G^{(N)}$ is real. The contribution to the Cauchy
integral comes from the pole $\delta g =-g_*$ which tends to zero
as $t\to t_0.$ The values of functional integrals on the saddle
point configurations are proportional to the entire volume of
functional integration, they are cancelled in the numerator and
denominator simultaneously. While calculating the fluctuation
integral, we take into account that the first order contributions
in $\delta u$ and $\delta u'$ are absent because of the
saddle-point condition. We also neglect the high-order
interactions between fluctuations, $O(\delta \phi^3),$ $O(\delta
\phi^4),$ \textit{etc.} to arrive at the Gaussian functional
integral,
\begin{equation}\label{32}
  G^{(N)}(r,\delta t)\simeq_{\delta t\rightarrow 0}
  \frac{g_*^N N^{N+1/2}}{2\pi}
 \int\int \mathcal{D}\delta u{\ }\mathcal{D}\delta u'{\ }
\exp -\frac N2 \mathrm{Tr}\left[ \delta u'\Lambda{\ } \delta u
+\delta u\left( \frac 1r \partial_r\right)\delta u \right].
\end{equation}
Performing the usual rescaling of fluctuation fields
$$
\delta u \to \delta u /\sqrt{N}, \quad \delta u' \to \delta
u'/\sqrt{N},
$$
we compute the Gaussian integral, with respect to $\delta u$
first, and then the resulting Gaussian integral over $\delta u'$,
\begin{equation}\label{33}
G^{(N)}(r,\delta t)\simeq_{\delta t\rightarrow 0}N^{N-1/2}\exp
\left(-N \log g_* \right)\det \Lambda^{-1}
\left(1+O\left(\frac{1}{N}\right)\right) .
\end{equation}
 The kernel of operator $\Lambda^{-1}$ is the Green function
of the linear diffusion equation (\ref{14}). Using the Stirling's
formula, one can check that the coefficients $G^{(N)}$ of the
asymptotic series (\ref{13}) demonstrate the factorial grows (like
in the most of quantum field theory models ):
\begin{equation}\label{34}
G^{(N)}(r,\delta t)\simeq_{\delta t\rightarrow 0} \frac{N!}{2\pi
N}\exp N\left(1-\log g_*\right)\det \Lambda^{-1}
\left(1+O\left(\frac{1}{N}\right)\right) .
\end{equation}
Therefore, the asymptotic series (\ref{13}) can be summed by means
of Borel's procedure. It consists of the following transformation
of series (\ref{13})
\begin{equation}\label{35}
\sum_{N}G^{(N)}g^N=\sum_{N}\Gamma(N+1)\mathcal{G}^{(N)}g^{N}=\sum_{N}\int_{0}^{\infty}
d\tau {\ } \mathcal{G}^{(N)} (g\tau)^{N} e^{ -\tau}
\end{equation}
where $ \mathcal{G}^{(N)} =G^{(N)}/\Gamma(N+1)$ are the new
expansion coefficients which do not exhibit the factorial growth.
We change the orders of summation and integration in (\ref{35})
that is indeed incorrect from the mathematical point of view, but
is broadly used in physics,
\begin{equation}\label{36}
\sum_{N}\int_{0}^{\infty} d\tau {\ } \mathcal{G}^{(N)} (g\tau)^{N}
e^{ -\tau} = \int_{0}^{\infty} d\tau {\ }e^{ -\tau}\sum_{N}
\mathcal{G}^{(N)}(g\tau)^{N}.
\end{equation}
We sum over $N$ in the rhs of (\ref{36}),
\begin{equation}\label{37}
 \frac{\det \Lambda^{-1}}{2\pi}\int_{0}^{\infty}
 d\tau {\ }e^{ -\tau}\sum_{N=1}^{\infty}
 \frac{(g\tau)^{N}}{N}e^{N\left( 1- \log
 g_*\right)} =-\frac{\det \Lambda^{-1}}{2\pi}\int_{0}^{\infty}
 d\tau {\ }e^{ -\tau} \log\left(1-\tau\frac{g}{ g_*} \right),
\end{equation}
and integrate it over $\tau$,
\begin{equation}\label{38}
  G(g)\simeq _{\delta t \to 0}\det \Lambda^{-1}\left(
  1+\frac{1}{2\pi}
  \mathrm{Ei}\left( \frac{g_*}{g}\right) \exp \left(-\frac{g_*}{g}\right)\right)
\end{equation}
where $\mathrm{Ei}(x)$ is the exponential integral. The asymptotic
diffusion kernels accounting for the  corrections due to the
bifurcation processes,
\begin{equation}\label{39}
G(x - x_0 ,t - t_0  )\simeq _{t \to t_0} \frac{\exp \left(
-\frac{(x - x_0)^2}{4\nu ( t - t_0 )}\right)} {\left( 4\pi \nu (t-
t_0)\right)^{d/2}}\left[
  1+\frac{1}{2\pi}
  \mathrm{Ei}\left( \frac{t - t_{0}}{\left( {x - x_{0}}
   \right)^{2}}\right) \exp \left(-\frac{t - t_{0}}{\left( {x - x_{0}}
   \right)^{2}}\right)\right],
\end{equation}
are sketched on Fig.~6. by the dash-dot lines. The unperturbed
diffusion kernels are shown by the solid lines.

\section{Discussion and Conclusion}
\noindent

In the present paper, we have studied the hydrodynamical flows of
incompressible fluids as the Brownian motions over the space of
fluid velocity configurations. The essential point of our approach
is that we have used the field theory formulation of the dynamics
which allowed us for the implementation of various powerful
technics borrowed from the quantum field theory. In particular, we
have investigated the Green function for the nonlinear
hydrodynamical equations.

In the framework of proposed approach, the Green functions can be
calculated in the form of well defined asymptotic series. The
numerical simulations have demonstrated that these series converge
relatively fast. We have studied the high order coefficients of
the asymptotic series around the unperturbed diffusion kernel
describing the pure relaxation dynamics. We have shown that
similarly to the most of quantum field theory models the high
order contributions into the Green function exhibit a factorial
growth. We have summed over the asymptotic series by means of the
standard Borel summation procedure and found the closed analytical
form for the corrections (up to an infinite order) to the pure
diffusion kernel.

The value of our paper is that it provides a possible ground for
the optimization of existing numerical simulation algorithms for
the large scale simulations in hydrodynamics. In particular, it
can be used in studies of the cross-field turbulent transport in
plasmas. Moreover, we have developed the method that can be
successfully used in the turbulence studies. It is worth to
mention that in the present paper we have constructed the
asymptotic solution for the Navier-Stocks equation in the
"vicinity" of standard diffusion kernel. However, other
asymptotical regimes would also be possible and have to be
studied.

\section{Acknowledgements}
 \noindent

One of the authors (D.V.)  is grateful to the Alexander von
Humboldt Foundation (Germany) and C.N.R.S. de France, Delegation
Provence, for the support that he gratefully acknowledges. While
preparing the paper, D.V. benefited from the hospitality of the
Zentrum fur Interdisciplinaere Forschung, Universitaet Bielefeld
(Germany).

\newpage

\begin{figure}[hp]
 \noindent
 \begin{minipage}[b]{.36\linewidth}
 \begin{center}
 \epsfig{file=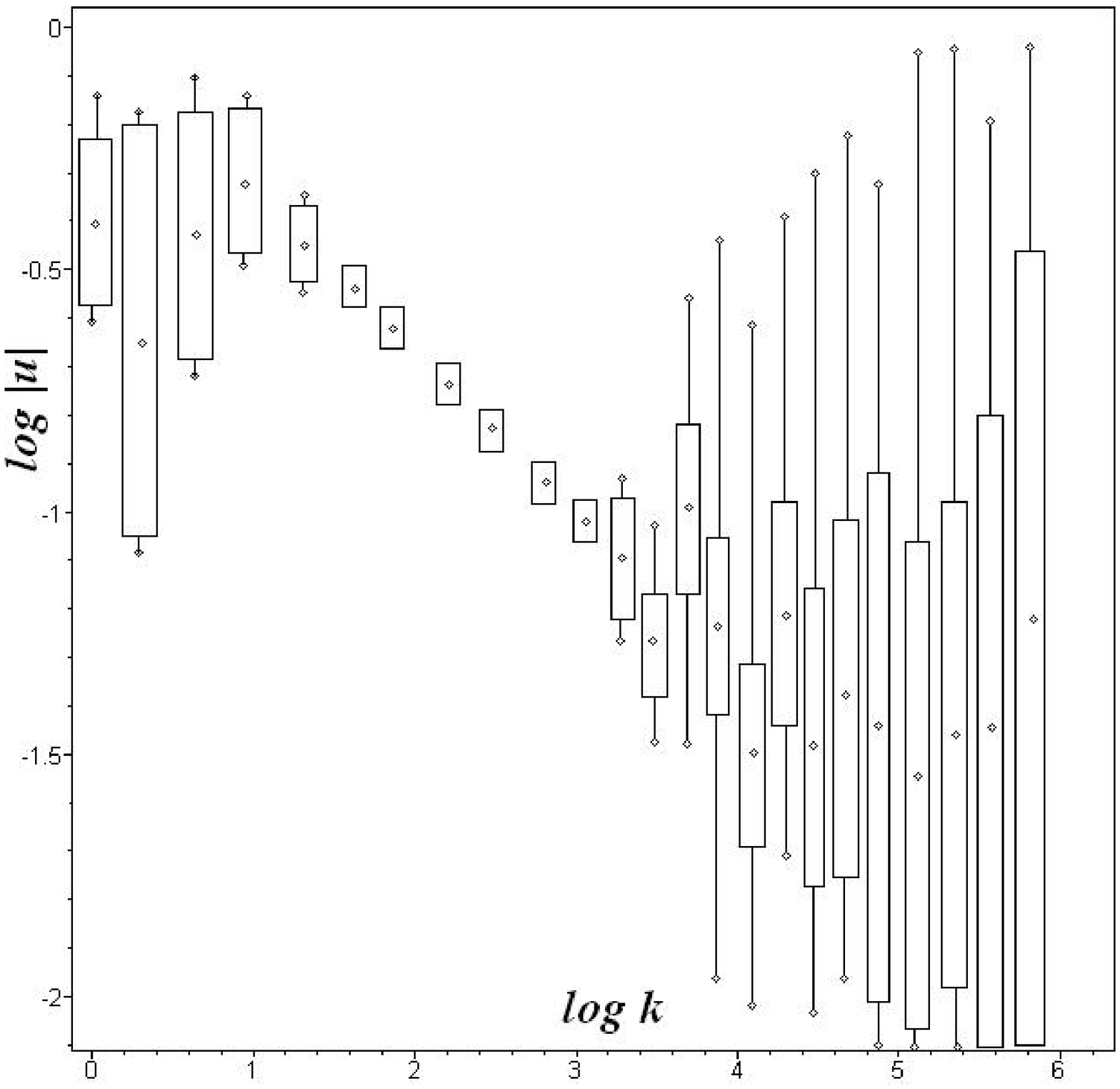, angle= 0,width =15cm, height =13cm}
 \end{center}
\end{minipage}
\label{F1}
\caption{ The expectations of the multiplicative binary
branching process for the 3D FNS forced in the small moments
$k=O(1)$ (in the large scales). The binary trees have been sampled
independently for each velocity component, but  the total number
of samples is limited just to 250 for each point $(k,t)$. The set
of randomly generated 250 samples obviously does not provide a
reliable base for the turbulent statistics for the large $k$ that
is evident from the excessively strong trembling of data worsening
as the momentum $k$ grows up. }
\end{figure}

\begin{figure}[hp]
 \noindent
 \begin{minipage}[b]{.36\linewidth}
 \begin{center}
 \epsfig{file=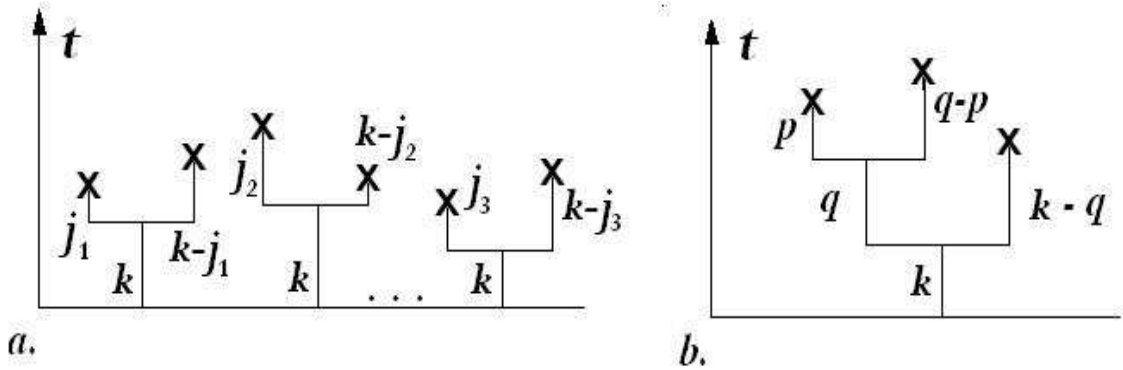, angle= 0,width =15cm, height =5cm}
 \end{center}
\end{minipage}
\label{F2} \caption{ a). The statistical ensemble of the pitchfork
bifurcations relevant to the set of possible branching random
walks. The branching random walk holds in state $k$ for an
exponential time characterized by the parameter $\nu k^{2}$. When
clock rings, the vortex bifurcate into two offspring with moments
$j \in {\left[ {1,k - 1} \right]}$ and $k - j$. They exist for the
exponential times accordingly to their moments and then dissipate
that is figured by the crosses. b). The branching random walks
with the double pitchfork bifurcations.
 }
\end{figure}

\begin{figure}[hp]
 \noindent
 \begin{minipage}[b]{.36\linewidth}
 \begin{center}
 \epsfig{file=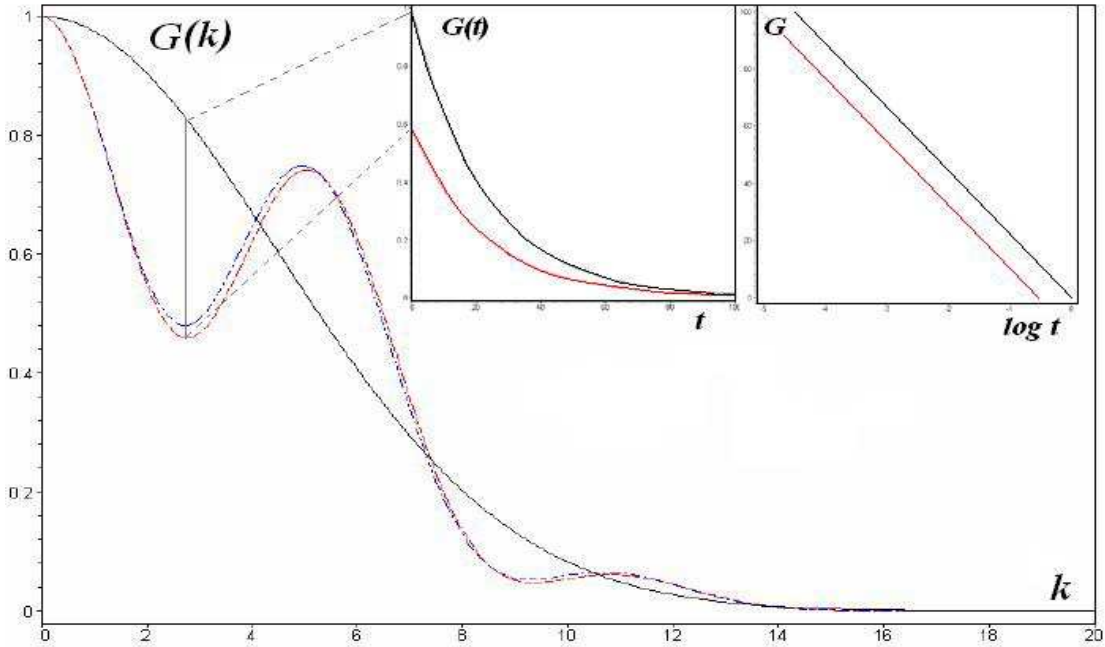,  angle= 0,width =15cm, height =9cm}
 \end{center}
\end{minipage}
\label{F3} \caption{ The $k$-spectrum of the unperturbed 1D
diffusion kernel at $t = 5,$ $\nu = 0.005$ is shown by the solid
line. The dash line presents the $G\left( {k} \right)$-spectrum
accounting for the corrections due to the pitchfork bifurcations.
The $G\left( {k} \right)$- spectrum accounting for the corrections
due to both the second and the third diagrams in (\ref{13}) is
displayed on Fig.~3 by the dash-dot line. On the outlined graphs,
we have displayed the exponential decay of $G$-spectrums at given
momentum $k$.}
\end{figure}

\begin{figure}[hp]
 \noindent
 \begin{minipage}[b]{.36\linewidth}
 \begin{center}
 \epsfig{file=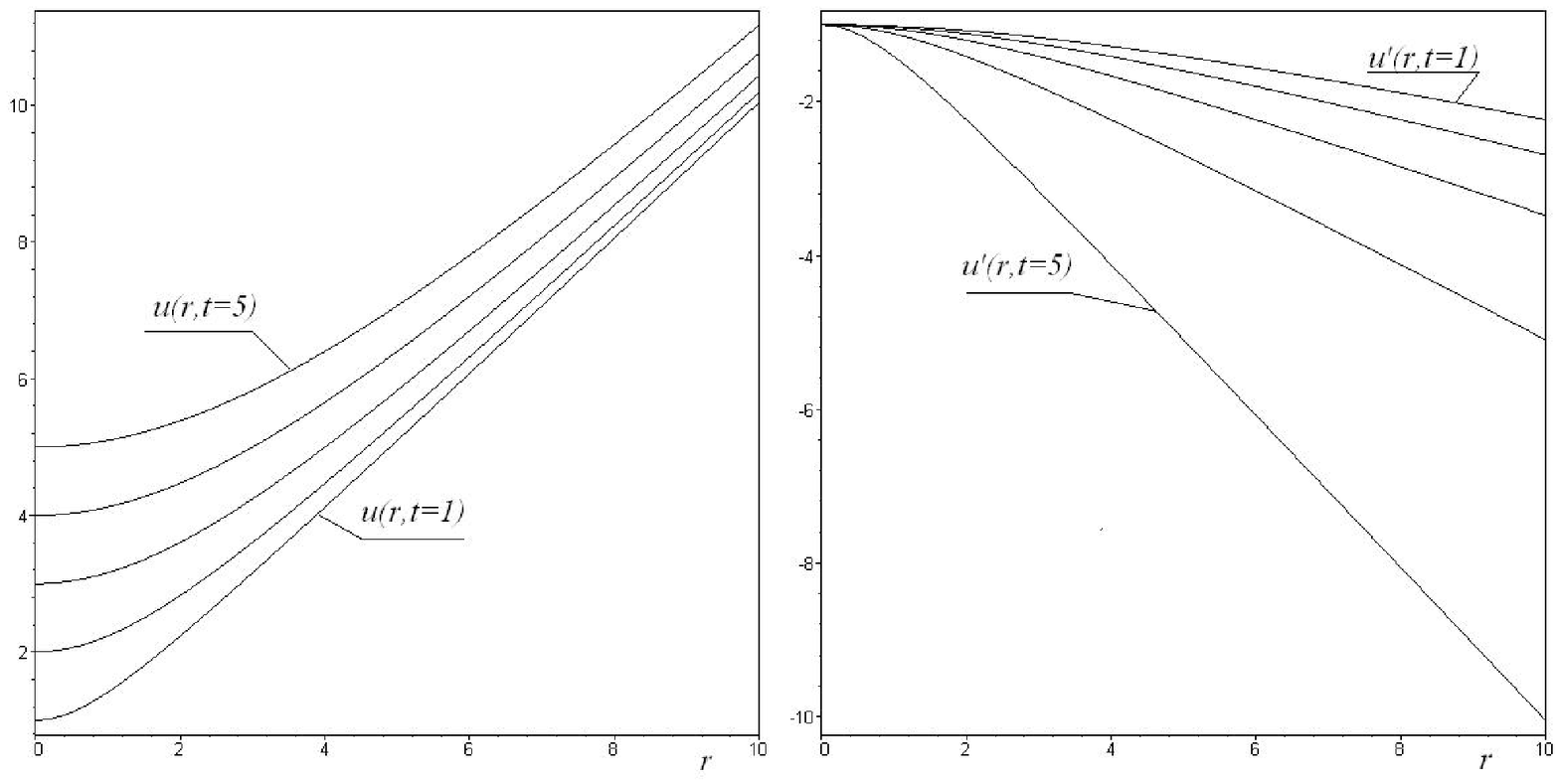,  angle= 0,width =15cm, height =9cm}
 \end{center}
\end{minipage}
\label{F4} \caption{The radial profiles (for 5 consequent time
steps) of the radially symmetric velocity field $u(r,t)$ and the
auxiliary field $u'(r,t)$ which meet the saddle-point equations
with respect to the high order asymptotic contributions into the
Green function. These solutions correspond to the effective value
$g_*\to 0$ of the coupling constant in the hydrodynamical equation
and contribute the asymptotic close to the standard diffusion
kernel. }
\end{figure}

\begin{figure}[hp]
 \noindent
 \begin{minipage}[b]{.36\linewidth}
 \begin{center}
 \epsfig{file=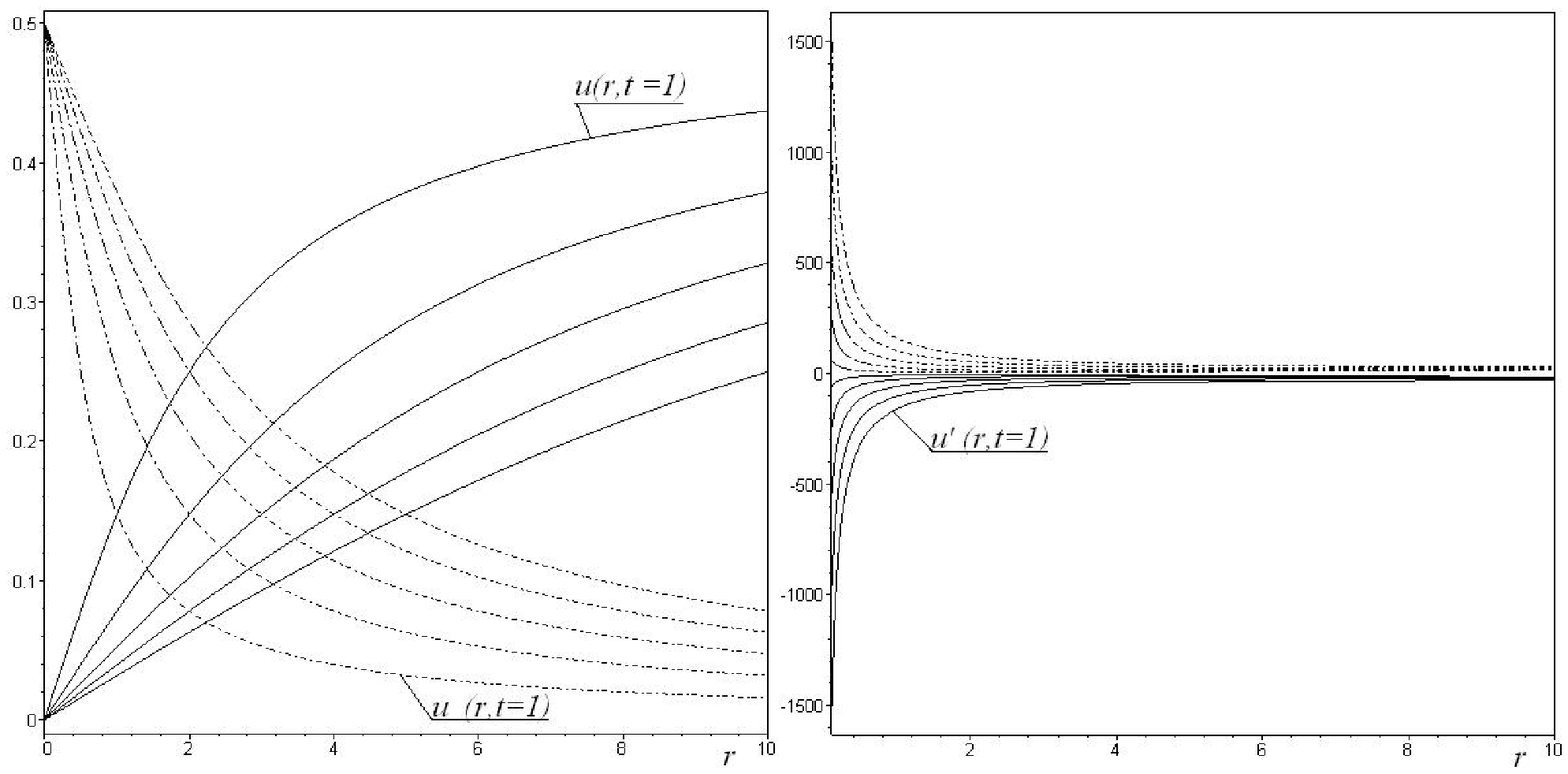,  angle= 0,width =15cm, height =9cm}
 \end{center}
\end{minipage}
\label{F5} \caption{The radial profiles (for 5 consequent time
steps) of the radially symmetric velocity field $u(r,t)$ and the
auxiliary field $u'(r,t)$ which meet the saddle-point equations
with respect to the high order asymptotic contributions into the
Green function. The solutions correspond to the effective value
$g_*\to \infty$ of the coupling constant in the hydrodynamical
equation. They do not contribute into the high order asymptotic
coefficients.}
\end{figure}

\begin{figure}[hp]
 \noindent
 \begin{minipage}[b]{.36\linewidth}
 \begin{center}
 \epsfig{file=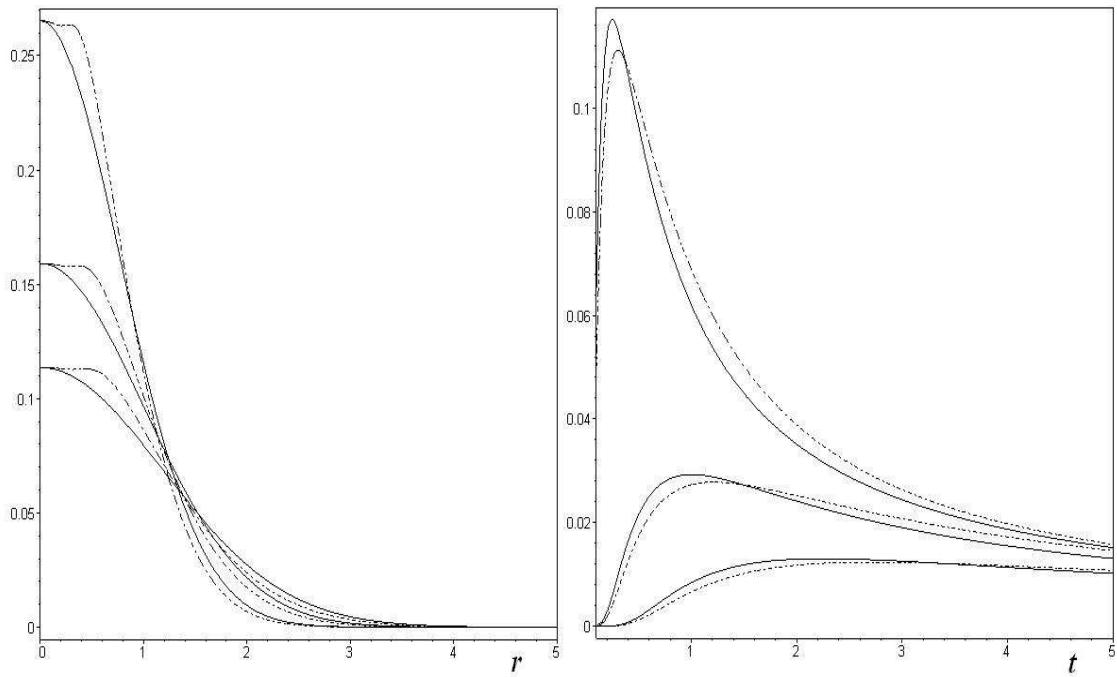,  angle= 0,width =15cm, height =9cm}
 \end{center}
\end{minipage}
\label{F6} \caption{ The profiles of standard diffusion kernel
(the solid line). On the left: $G(r)$ at several consequent time
steps; on the right: $G(t)$ at several distant points. The
dash-dot lines present the asymptotic kernel (as $t\to t_0$)
accounting for all bifurcations (up to an infinite order) of
vortexes. }
\end{figure}
\end{document}